\documentclass[sigconf]{acmart}

\usepackage{booktabs}

\usepackage{enumerate}
\usepackage{epsfig}
\usepackage{graphicx}
\usepackage{amsmath}
\usepackage{amssymb}
\usepackage{algorithm}
\usepackage{url}
\usepackage{listings}
\usepackage{amsmath}
\usepackage{rotating}
\usepackage{wrapfig}
\usepackage{multirow}
\usepackage{lscape}
\usepackage{verbatim}
\usepackage{wrapfig}
\usepackage{hhline}
\usepackage{textcomp,booktabs}
\usepackage{xcolor}
\usepackage{paralist}
\usepackage{caption}
\DeclareCaptionType{copyrightbox}
\usepackage{subcaption}
\usepackage{grffile}
\usepackage{gensymb}
\usepackage{flushend}

\newtheorem{claim}{Claim}

\copyrightyear{2017}
\acmYear{2017}
\setcopyright{acmcopyright}
\acmConference{WebSci'17}{}{June 25--28, 2017, Troy, NY, USA}
\acmPrice{15.00}
\acmDOI{http://dx.doi.org/10.1145/3091478.3091507}
\acmISBN{978-1-4503-4896-6/17/06}

\begin{document}

\title{Stateless Puzzles for Real Time Online Fraud Preemption}

\author{Mizanur Rahman}
\affiliation{
   \institution{Florida Int'l University, USA}
}
\email{mrahm031@fiu.edu}
\author{Ruben Recabarren}
\affiliation{
   \institution{Florida Int'l University, USA}
}
\email{recabarren@gmail.com}
\author{Bogdan Carbunar}
\affiliation{
   \institution{Florida Int'l University, USA}
}
\email{carbunar@gmail.com}
\author{Dongwon Lee}
\affiliation{
   \institution{Penn State University, USA}
}
\email{dongwon@psu.edu}

\begin{abstract}
The profitability of fraud in online systems such as app markets and social
networks marks the failure of existing defense mechanisms. In this paper, we
propose {\bf FraudSys}, a real-time fraud preemption approach that imposes
Bitcoin-inspired computational puzzles on the devices that post online system
activities, such as reviews and likes. We introduce and leverage several novel
concepts that include (i) stateless, verifiable computational puzzles, that
impose minimal performance overhead, but enable the efficient verification of
their authenticity, (ii) a real-time, graph based solution to assign fraud
scores to user activities, and (iii) mechanisms to dynamically adjust puzzle
difficulty levels based on fraud scores and the computational capabilities of
devices.  FraudSys does not alter the experience of users in  online systems,
but delays fraudulent actions and consumes significant computational resources
of the fraudsters.  Using real datasets from Google Play and Facebook, we
demonstrate the feasibility of FraudSys by showing that the devices of honest
users are minimally impacted, while fraudster controlled devices receive daily
computational penalties of up to 3,079 hours.  In addition, we show that with
FraudSys, fraud does not pay off, as a user equipped with mining hardware
(e.g., AntMiner S7) will earn less than half through fraud than from honest
Bitcoin mining.
\end{abstract}

%\begin{CCSXML}
%<ccs2012>
%<concept>
%<concept_id>10002978.10003022.10003026</concept_id>
%<concept_desc>Security and privacy~Web application security</concept_desc>
%<concept_significance>500</concept_significance>
%</concept>
%<concept>
%<concept_id>10002978.10003022.10003027</concept_id>
%<concept_desc>Security and privacy~Social network security and privacy</concept_desc>
%<concept_significance>500</concept_significance>
%</concept>
%<concept>
%<concept_id>10010147.10010257.10010258.10010259.10010263</concept_id>
%<concept_desc>Computing methodologies~Supervised learning by classification</concept_desc>
%<concept_significance>300</concept_significance>
%</concept>
%</ccs2012>
%\end{CCSXML}
%
%\ccsdesc[500]{Security and privacy~Web application security}
%\ccsdesc[500]{Security and privacy~Social network security and privacy}
%\ccsdesc[300]{Computing methodologies~Supervised learning by classification}


\maketitle

\section{Introduction}

The social impact of online services built on information posted by their users
has also turned them into a lucrative medium for fraudulently influencing
public opinion~\cite{RRCC16,BLLTZ16,LZ16,SWEKVZZ13}.  The need to aggressively
promote disinformation has created a black market for social network fraud,
that includes fake opinions and reviews, likes, followers and app
installs~\cite{MMLSV11,TSM,RR,RL,AV,AS,AR}. For instance, in
$\S$~\ref{sec:model:adversary}, we show that in fraud markets, a fake review
can cost between \$0.5 and \$3 and a fake social networking ``like'' can cost
\$2.  The profitability of fraud suggests that current solutions that focus on
fraud detection, are unable to control organized fraud.

%For instance, Figure~\ref{fig:fraudsites} shows the average, minimum and
%maximum price requested to purchase a single fake review in several
%crowdsourcing and fraud-as-a-service (FAAS) sites. Depending on the quality of
%the review and the quantity of purchase, a fake review can cost between \$0.5
%and \$3. The profitability of fraud suggests that current solutions that focus
%on fraud detection, are unable to control organized fraud.

\begin{figure}
\centering
\includegraphics[width=0.47\textwidth]{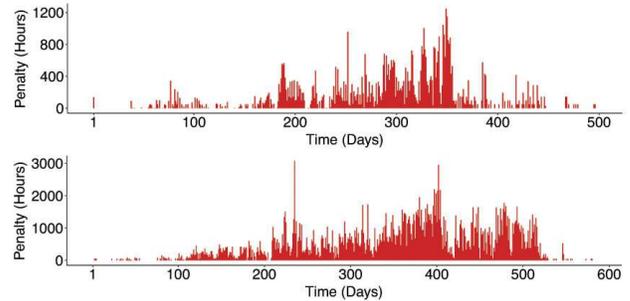}
%\vspace{-5pt}
\caption{Timeline of daily penalties (in hours) assigned by FraudSys to the
Google Play activities of two fraudsters we identified in Freelancer.com.
%
%when $minf$ = 1h and $maxf$ = 24h.
%
FraudSys imposes daily penalties of up to 1,247 hours to the fraudster at the
top and 3,079 hours for the fraudster at the bottom. As a result, the
fraudsters need to consume significant computational resources, while their
fraud is significantly delayed. This in turn reduces the number of payments
they would receive, and impacts their profitability.}
\label{fig:timeline}
\vspace{-15pt}
\end{figure}

In this paper we introduce the concept of {\it fraud preemption systems},
solutions deployed to defend online systems such as social networks and app
markets. Instead of reacting to fraud posted in the past, fraud preemption
systems seek to discourage fraudsters from posting fraud in the first place. We
propose FraudSys, the first real-time fraud preemption system that reduces the
profitability of fraud from the perspective of both crowdsourced fraud workers
and the people who hire them. FraudSys imposes computational penalties: the
activity of a user (e.g., review, like) is posted online only after his device
solves a computational puzzle. Puzzles reduce the profitability of fraud by (i)
limiting the amount of fraud per time unit that can be posted for any subject
hosted on the online system, and (ii) by consuming the computational resources
of fraudsters. For instance, Figure~\ref{fig:timeline} shows the timelines of
daily penalties assigned by FraudSys to two fraudsters detected in Google
Play. Based only on the recorded activities, FraudSys frequently assigned
hundreds of hours of daily computational penalties to a single fraudster.

\vspace{0.05in}\noindent
{\bf Challenges}.
Implementing a fraud preemption system raises several challenges.  First,
FraudSys needs to detect fraud in real-time, whenever a user performs an
online system activity. Once assigned, a puzzle cannot be rescinded.
This is in contrast to existing systems (e.g., Yelp) that detect fraud
retroactively and can update previous decisions when new information surfaces.
Second, FraudSys needs to impose difficult puzzles on fraudsters, but minimally
impact the experience of honest users.  This is made even more complex by the
fact that fraudsters can attempt to bypass detection and even obscure their
true ability to solve puzzles. Third, a stateful FraudSys service that
maintains state for millions of issued and active puzzles is expensive and
vulnerable to DoS attacks.

\vspace{0.05in}\noindent
\noindent
{\bf Our Contributions}.
Through FraudSys, we introduce several innovative solutions. To address the
first challenge, we exploit observations of fraudulent behaviors gleaned from
crowdsourcing sites and online systems, to propose a real-time graph based
algorithm to infer an {\it activity fraud score}, the chance that a user
activity is fraudulent [$\S$~\ref{sec:fraudsys:fraud}]. More specifically, we
introduce features that group fraudulent activities according to their human
creator: FraudSys identifies densely connected components in the co-review
graph of the subject targeted by the user activity, each presumably controlled
by a different fraudster. It then quantifies the connectivity of the user
account performing the action, to each component, and uses the highest
connectivity as features that may indicate that the user account and the
corresponding component are controlled by the same fraudster. FraudSys then
leverages supervised learning algorithms trained on these features to infer the
activity fraud score.

To address the second challenge, we develop adaptive hashrate inference
techniques to detect the computational capabilities of even adversarial
controlled devices to solve puzzles [$\S$~\ref{sec:fraudsys:all}], and devise
mechanisms to convert fraud scores to appropriate temporal penalty and puzzle
difficulty values [$\S$~\ref{sec:fraudsys:all}]. The puzzles assigned by
FraudSys do not alter the online experience of users, as they are solved on
their devices, in the background. However, the puzzles (1) significantly delay
detected fraudulent activities, posted only when the device returns the correct
puzzle solutions and (2) consume the computational resources of the fraudsters
who control the devices.

To address the third challenge, we propose the notion of {\it stateless
computational puzzles}, computational tasks that impose no storage overhead on
the fraud preemption system provider, but enable it to efficiently verify their
authenticity and the correctness of their solutions
[$\S$~\ref{sec:fraudsys:puzzler}]. Thus, the fraud preemption system can assign
a puzzle to a device from which an activity was performed on the online system,
without storing any state about this task. The device can return the results of
the puzzle in 5 seconds or 1 day, and the provider can verify that the task is
authentic, and its results are correct. This makes our approach resistant to
DoS attacks that attempt to exhaust the provider's storage space for assigned
puzzles.

We show that the computational penalty imposed by FraudSys on a fraudulent
activity is a function of the capabilities of the device from which it is
performed, and the probability that the activity is fraudulent. We introduce
and prove upper bounds on the profitability of fraud and the amount of fraud
that can be created for a single subject, per time unit
[$\S$~\ref{sec:properties}] .
We evaluate FraudSys on 23,028 fraudulent reviews (posted by 23 fraudsters from
2,664 user accounts they control), and 1,061 honest reviews we collected from
Google Play, as well as 274,297 fake and 180,400 honest likes from Facebook.
Even with incomplete data, FraudSys imposes temporal penalties that can be as
high as 3,079 hours per day for a single fraudster.
We also show that fraud does not pay off. At today's fraud payout, a
fraudster equipped with an AntMiner S7 (Bitcoin mining hardware) will earn
through fraud less than half the payout of honest Bitcoin mining.

%\noindent
%FraudSys offers an adversary two options, to either return the puzzle solutions
%after the time penalty expires, or earlier. In the first case the rate and
%profitability of fraud are reduced. In the second case, FraudSys obtains a more
%accurate estimate of the capabilities of the device controlled by the
%adversary. The next puzzle assigned to this adversary will thus require more
%work, adding CPU to the list of resources consumed by the adversary.

%\noindent
%We emphasize that FraudSys can be flexibly implemented either as a component of
%the online service or as a third party service. 

%\vspace{-20pt}

\section{Related Work}

\noindent
{\bf Computation Based Fraud Preemption}.
Dwork and Naor~\cite{DN92} were the first to propose the use of computation to
prevent fraud, in particular spam, where the sender of an e-mail needs to
include the solution to a ``moderately hard function'' computed over a function
of the e-mail.  Juels and Brainard~\cite{JB99} proposed to use puzzles to
prevent denial of service attacks, while Borisov~\cite{B06} introduced puzzles
that deter Sybils in peer-to-peer networks. In Borisov~\cite{B06}, newly joined
peers need to solve a puzzle to which all the other peers have contributed.

FraudSys not only seeks to adapt computational puzzles to prevent online system
fraud, but also needs to solve the additional challenges of building puzzles
whose difficulty is a function of the probability that an activity is
fraudulent, while handling heterogeneous user devices (e.g., ranging from
smartphones to machines that specialize in such puzzles).

%FraudSys also needs to minimally impact the experience of honest users.

%\noindent
%{\bf Similarities to recaptcha}.
%FraudSys is similar to Turing test to deter spam users in recaptcha.
%In addition, legit users who pass a turing test can
%register a new account in recaptcha while legit users who solves a bitcoin
%piece can post SN activities such as reviews or Likes. Also, a recognized OCR
%by legit users in recaptcha rewards OCR researchers, while a solved bitcoin
%block rewards the service providers.

\vspace{0.05in}\noindent
{\bf Graph Based Fraud Detection}.
Graphs have been used extensively to model relationships and detect fraudulent
behaviors in online systems. Ye and Akoglu~\cite{YL15} quantified the chance of
a subject to be a spam campaign target, then clustered spammers on a 2-hop
subgraph induced by the subjects with the highest chance values.  Lu et
al.~\cite{LZXL13} proposed a belief propagation approach implemented on a
review-to-reviewer graph, that simultaneously detects fake reviews and
spammers (fraudsters).

%Van Vlasselaer et al.~\cite{VAESB15} detected
%fraudulent companies by identifying cliques in bipartite company-to-resource
%graphs, and by propagating ``exposure scores'' through the graphs.

Mukherjee et al.~\cite{MLG12} proposed a suite of features to identify reviewer
groups, as users who review many subjects in common but not much else, post
their reviews within small time windows, and are among the first to review the
subject.  Hooi et al.~\cite{HSBSSF16} have recently shown that fraudsters have
evolved to hide their traces, by adding spurious reviews to popular items. To
identify ``camouflaged'' fraud, Hooi et al.~\cite{HSBSSF16} introduced
``suspiciousness'' metrics that apply to bipartite user-to-item graphs, and
developed a greedy algorithm to find the subgraph with the highest
suspiciousness. Akoglu et al.~\cite{ATK15} survey graph based online
fraud detection. \cite{FH16} provide a survey of community detection methods,
evaluation scores and techniques for general networks.

%Zafarani and Liu~\cite{ZL15} attempt to detect malicious users using minimal
%information (i.e., mostly the user's name), in an effort to devise a solution
%that is applicable to most social networks. This makes it suitable to detect
%malicious users at registration, as users need to provide at least their
%username.  Instead of malicious users, FraudSys seeks to detect fraudulent
%actions, then penalizes them according to their likelihood of being
%fraudulent.

Unlike previous work, FraudSys assigns fraud scores to individual user
activities in {\it real time}, thus uses only partial information. To achieve
this, FraudSys develops and leverages features that quantify the connectivity
of the user activity to other groups of activities {\it previously} performed
by other fraudsters on the same subject. Further, FraudSys also imposes
computation and temporal penalties to discourage fraud creation.

\section{System and Adversary Model}
\label{sec:model}

\begin{figure}
\centering
\includegraphics[width=0.47\textwidth]{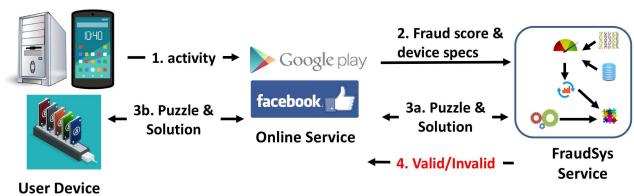}
\caption{System model. The user performs actions on the online service, from a
device that can range from a smartphone to a Bitcoin miner. The online service
implements and posts the activity only if and after the FraudSys service
validates it. The FraudSys functionality can be implemented by the online
service or by a third party provider.}
\label{fig:system}
\vspace{-15pt}
\end{figure}

Figure~\ref{fig:system} illustrates the three main components of the system
model. First, the online service (the \textit{service}) hosts the system
functionality, and stores information about user accounts and featured
\textit{subjects}. Subjects can be apps in stores like Google Play, or pages
for businesses, accounts and stories in social networks like Facebook.

Second, the users: they register with the service, record profile information
(e.g., name) and receive initial service credentials, including a unique id.
Users can access the online service from a variety of devices. For this, they
need to install a client (e.g., app) on each device they use.  The online
service stores and maintains information about each device that the user has
used, e.g., to provide compatibility information on Google Play apps.

%We detail the storage later in the paper.

Users are encouraged to {\it act} on existing subjects. The activities include
posting reviews, comments, or likes, installing mobile apps, etc.  The online
service associates statistics over the activities performed for each supported
subject. The statistics have a significant impact on the popularity and search
rank of subjects~\cite{Ankit.Jain,RankGooglePlay}, thus are targets of manipulation by
fraudsters (see $\S$~\ref{sec:model:adversary}).

%Subjects that rank higher in search results tend to be more popular, see
%e.g.,~\cite{RankGooglePlay}.

The third component of the system model is the FraudSys service, whose goal is
to validate user activities. For increased flexibility, Figure~\ref{fig:system}
shows FraudSys as an independent provider. However, FraudSys can also be a
component of the online service.

%This approach raises an additional challenge, that online service providers
%may lack the required puzzle creation expertise, and may be reluctant to mine
%Bitcoins with their user's resources.  To address this challenge, we propose
%that part of the FraudSys functionality is to be provided by a third party
%service.  The incentive for providing this service is financial: FraudSys can
%provide this service for a multitude of online services concerned by fraud,
%e.g., social networks, app markets, location based services, and even security
%services such as CloudFlare.

\subsection{Adversary Model}
\label{sec:model:adversary}

We consider two types of adversaries -- adversarial owners and crowdsourced
fraud workers.

\vspace{0.05in}\noindent
{\bf Adversarial owners}.
Adversarial behaviors start with the subject owners. Adversarial owners seek to
fraudulently promote their subjects (or demote competitor subjects) in order to
bias the popularity and public opinion of specific subjects. For instance,
fraudulent promotions seek to make subjects more profitable~\cite{LZ16,AM12},
increase the ``reachability'' of malware (through more app installs), and boost
the impact of fake news.

\vspace{0.05in}\noindent
{\bf Fraud workers ($=$ fraudsters)}.
We assume that adversarial owners crowdsource this promotion task (also known
as search rank fraud) to {\it fraud workers}, or fraudsters. In this paper we
focus on two types of fraudulent activities: writing fake reviews in Google
Play and posting fake ``Likes'' in Facebook.  We have studied fraudster
recruitment jobs in crowdsourcing sites and fraud posted in Google Play and
Facebook. This has allowed us to collect fraud data (see $\S$~\ref{sec:data})
and to identify several fraud behaviors: (i) more than one fraudster can target
the same subject; (ii) user accounts controlled by a fraudster tend to have a
significant history of common activities, i.e., performed on the same subjects;
and (iii) accounts controlled by different fraudsters tend to have few common
past activities.

\begin{figure}
\centering
\includegraphics[width=0.47\textwidth]{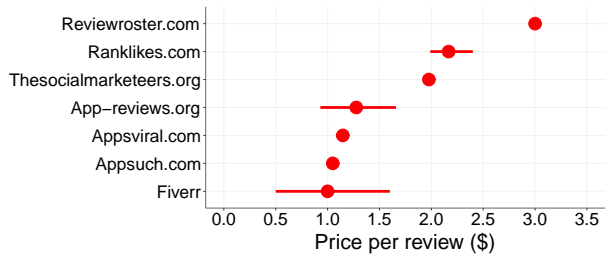}
\caption{Price per review (minimum, average and maximum), for crowdsourcing
sites that focus on app market fraud. The sites offer ``fraud packages'' and
even discounts for bulk fake review purchases. A fake review costs between
\$0.5-\$3.
\label{fig:fraudsites}}
\vspace{-15pt}
\end{figure}

%The fraud detection module of FraudSys exploits these observations to assign
%fraud scores to user activities.

%In addition to general crowdsourcing sites such as
%Freelancer~\cite{Freelancer}, Upwork~\cite{Upwork} and Fiverr~\cite{Fiverr}
%that host fraud workers, specialized ``fraud-as-a-service'' (FAAS) sites
%exist, where clients buy fraud packages without direct interaction with the
%workers~\cite{TSM,RR,RL,AV,AS,AR}. Figure~\ref{fig:fraudsites} shows the
%minimum, average and maximum cost per fraudulent activity, as advertised by
%several FAAS sites: a fake review for an app can worth as low as \$1, while a
%fake social networking ``like'' can cost as low as \$2.

\vspace{0.05in}\noindent
{\bf Fraud incentives}.
We assume that fraud workers are rational, motivated by financial incentives.
That is, given an original investment in expertise and equipment, a fraud
worker seeks to maximize his revenue achieved per time unit.
Figure~\ref{fig:fraudsites} shows the minimum, average and maximum cost per
fraudulent activity, as advertised by several crowdsourcing and
fraud-as-a-service (FAAS) sites: a fake review for an app is worth between
\$0.5-\$3, while a fake social networking ``like'' can cost \$2.  In contrast,
an adversarial owner may have both financial incentives (e.g., increased
market share for his subject, thus revenue), and external incentives (e.g.,
malware or fake news distribution).

\subsection{Fraud Preemption System Definition}
\label{sec:model:problem}

We introduce the concept of {\it fraud preemption systems}, that seek to
restrict the profitability of fraud for both fraudsters and the people who hire
the fraudsters (i.e., adversarial owners). Specifically, let $Sys =
(\mathcal{U}, \mathcal{S}, \mathcal{F}, P)$ be a system that consists of finite
sets of users ($\mathcal{U}$), subjects ($\mathcal{S}$) and fraudsters
($\mathcal{F}$) that interact through a set of procedures $P$. In the adversary
model of $\S$~\ref{sec:model:adversary}, we say that $Sys$ is a (p,a)-{\it
fraud preemption system} if it satisfies the following two conditions:

\begin{enumerate}

\item
{\bf Fraudster deterrence}:
The average payout per time unit of any fraudster in $\mathcal{F}$ does not
exceed $p$.

\item
{\bf Adversarial owner deterrence}:
The average number of fraudulent activities allowed for any subject in
$\mathcal{S}$ per time unit does not exceed $a$.

\end{enumerate}

\noindent
In addition, a puzzle-based fraud preemption system needs to satisfy the
following requirements:

\begin{enumerate}

\item
{\bf Real-time fraud detection}.
Detect fraud at the time it is created, with access to only limited information
(i.e., no knowledge of the future).

\item
{\bf Penalty accuracy}.
Impose difficult puzzles on fraudsters, but minimally impact the online
experience of honest users.

\item
{\bf Device heterogeneity}.
Both honest and fraudulent users may register and use multiple devices to
access the online service. Malicious users may obfuscate the computational
capabilities of their devices.

\item
{\bf Minimize system resource consumption}.
The high number of issued, active puzzles will consume the resources of the
FraudSys provider, and open it to DoS attacks.

\end{enumerate}

%\begin{compactitem}
%
%\item
%{\bf Delay fraud impact}.
%Impose delays on the time taken by posted fraud to have an impact.
%
%\item
%{\bf Curb fraud per account}.
%Limit the amount of fraud that fraud worker can create per time unit from a
%single user account.
%
%\item
%{\bf Minimize impact on honesty}.
%Minimally interfere with the activities of honest users.
%
%\end{compactitem}

%The goal of the adversary is to make financial profit. Specifically, let $P$
%be the daily profit made by an adversary from a single device. Then, $P = E -
%C$, where $E$ is the adversary's daily fraud job earnings, and $C$ is his
%daily cost for performing the fraud.  The cost $C$ includes the amortized cost
%of electricity, internet, and device depreciation. \bogdan{None of these costs
%are high. Electricity cost to charge a smartphone for a year is 25 cents,
%internet can be shared among many devices and uses, and the adversary can use
%very old and cheap devices.}

\section{FraudSys}
\label{sec:fraudsys}

\begin{figure}
\centering
\includegraphics[width=0.49\textwidth]{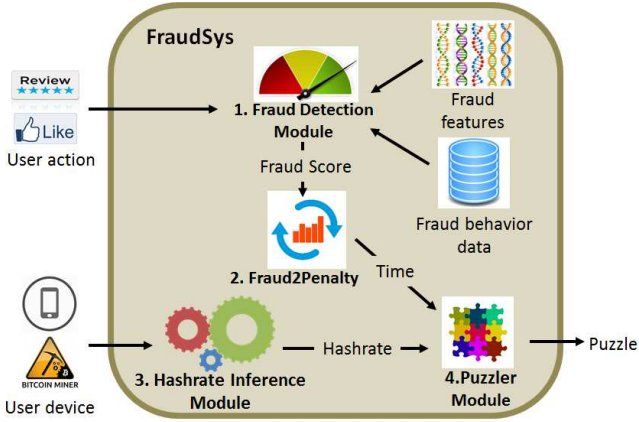}
%\vspace{5pt}
\caption{FraudSys architecture. The Fraud Detector module uses supervised
learning to assign a fraud score to user activities. The Fraud2Penalty module
converts the fraud score to a {\it time penalty}. The Hashrate Inference module
estimates the computational capabilities of the user device. Finally, the
Puzzler module generates a puzzle that the device should take approximately the
time penalty to solve.
\label{fig:fraudsys}}
\vspace{-15pt}
\end{figure}

We introduce FraudSys, a real-time fraud preemption system that requires users
to verify commitment through an imposed resource consumption action for each
activity they perform on the online system. Specifically,
FraudSys requires the device from which the activity was issued, to solve a
{\it computational puzzle}. FraudSys consists of the modules illustrated in
Figure~\ref{fig:fraudsys}: The Fraud Detection module takes as input a user
activity and the current state of the subject, and outputs a {\it fraud score}.
The Fraud2Penalty module converts the fraud score to a {\it time penalty}: the
time that the user's device will need to spend working on a computational
puzzle. The Hashrate Inference module interacts with the user device in order
to learn its puzzle solving capabilities.  Finally, the Puzzler module uses the
inferred device capabilities to generate a puzzle that the device will take
a time approximately equal to {\it time penalty} to solve.

To address requirement \#1, the Fraud Detection module exploits the fraudulent
behaviors described in $\S$~\ref{sec:model:adversary}. It builds {\it
co-activity graphs} and extracts features that model the relationships between
the user performing the activity and other users that have earlier performed
similar activities for the same subject.

We address requirement \#2 through a two-pronged approach. First, the Fraud
Detection and Fraud2Penalty modules ensure that the difficulty of a FraudSys
puzzle will be a function of the detected probability of fraud: activities
believed to be honest will be assigned trivial puzzles, while increasingly
fraudulent activities will be assigned increasingly difficult puzzles.  Second,
FraudSys does not change the experience of the user on the online system: the
user writes the review or clicks on the like button, then continues browsing or
quits the app. The assigned puzzle is solved in the background by the device on
which the activity was performed. However, FraudSys delays the publication of
the activity, until the device produces the correct puzzle solution.

To address requirement \#3, the Hashrate Inference module estimates the
hashrate of the device performing the activity, and provides the tool to punish
devices that cheat about their puzzle solving capabilities. To solve
requirement \#4, the Puzzler module generates puzzles that outsource the
storage constraints from the FraudSys service to the user devices that solve
the puzzles. In the following we detail each FraudSys module, starting with the
central puzzle creation module.

\begin{table}
\setlength{\tabcolsep}{.16677em}
\centering
\textsf{
\small
\begin{tabular}{l r}
\toprule
\textbf{Notation} & \textbf{Definition}\\
\midrule
$U$, $D$, $S$, $A$ & user, device, subject, activity\\
T & time of puzzle issue\\
$r$ & activity fraud score\\
\midrule
$\Delta$ & puzzle difficulty\\
$\eta_D$ & hashrate of device $D$\\
$\Gamma$ & puzzle cookie\\
$\Pi$ & puzzle\\
$target$ & puzzle target value\\
$\tau$ & temporal penalty\\
$q$ & number of shares (puzzle solutions)\\
\midrule
$K$ & secret key of FraudSys\\
\bottomrule
\end{tabular}
}
\caption{FraudSys symbol table.}
\label{fig:symbols}
\vspace{-15pt}
\end{table}

\subsection{The Puzzler Module: Stateless Puzzles}
\label{sec:fraudsys:puzzler}

Let $U$ be a user that performs an activity $A$ from a device $D$, on a subject
$S$ hosted by the online service.  Table~\ref{fig:symbols}
summarizes the notations we use.  The FraudSys service stores minimal state for
each registered user, and {\it serializes} his activities, see
Figure~\ref{fig:serialization}: the devices from which a user performs a
sequence of activities on the online service, are assigned one puzzle per
activity, each with its own timeout. The device needs to return the puzzle
solutions before the associated timeout. To implement this, for
each user $U$, the FraudSys service stores the following entry:
\[
U, [\langle D_i, \eta_i \rangle]_{i=1..d}, timeout,
\]
where, for each of the $i = 1 .. d$ devices registered by $U$, $D_i$ is the
device identifier and $\eta_i$ is its hashrate (puzzle solving capabilities
measure, see following), and $timeout$ is the latest time by which one of these
devices needs to return puzzle solutions.

FraudSys builds on the computational puzzles of Bitcoin, see~\cite{N08}.  Let
$H^2(M)$ denote the double SHA-256 hash of a message $M$. Then, the FraudSys
puzzle issued to device $D$ consists of a $target$ value and a fixed string
$F$. We detail $F$ shortly. To solve the puzzle, $D$ needs to randomly choose
32 byte long $nonce$ values until it finds at least one that satisfies:

\vspace{-5pt}

\begin{equation}
H^2(nonce || F) < target
\label{eq:puzzle}
\end{equation}

That is, the double hash of the $nonce$ concatenated with $F$, needs to be
smaller than the $target$ value, another 32 byte long value. A smaller $target$
implies a harder puzzle. The largest $target$ acceptable by the system
is called $target\_1$, or {\it target of difficulty 1}.

Bitcoin has two drawbacks. First, the current difficulty of Bitcoin puzzles
requires computational capabilities that greatly exceed those of devices used
to access online services. Second, Bitcoin requires the network to maintain
state about issued puzzles. State storage exposes FraudSys to attacks, while
not storing state can enable adversaries to lower the difficulty of their
assigned puzzles. To address these problems we (i) change the $target\_1$
difficulty to allow trivial puzzles, and (ii) introduce {\it puzzle cookies},
special values that authenticate puzzles with minimal FraudSys state, see
following.

\vspace{0.05in}\noindent
{\bf Device hashrate and puzzle difficulty}.
We set the $target\_1$ value to be a 32 byte long value with one zero at the
beginning, e.g., $2^{255}-1$. In addition, the {\it hashrate} $\eta_D$ of a
device $D$ is a measure that describes the ability of the device to solve
puzzles. Since the puzzles need to be solved in a brute force approach, the
hashrate is measured in hashes per second.  A relevant concept is the notion of
\textit{difficulty}, denoted by $\Delta$, a measure of how difficult it is to
solve a puzzle whose input values hash below a given target.  Its relationship
to the above $target$ value is given by:
\begin{figure}
\centering
\includegraphics[width=0.43\textwidth,height=0.9in]{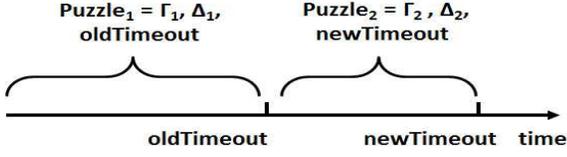}
\vspace{-5pt}
\caption{Puzzle serialization: a user can perform multiple activities, but each
receives a different puzzle with its own timeout, authenticated through the
cookie $\Gamma$.}
\label{fig:serialization}
\vspace{-10pt}
\end{figure}
\begin{equation}
\Delta = \frac{target\_1}{target} = \frac{2^{255}-1}{target}
\label{eq:delta:target}
\end{equation}
\noindent
Given $\eta_D$, we derive the time
$\tau$ taken by $D$ to solve a puzzle with difficulty $\Delta$, as follows.
First, the number of hashes smaller than a given target is equal to the target.
For instance, the number of hashes smaller than $target\_1$ is $2^{255}-1$ .
Then, the probability $p$ of finding an input that hashes to a value smaller
than the target is equal to the target divided by the total number of hashes
($2^{256}$). Furthermore, the expected number of hashes, $E$, before achieving
the target is given by $1/p$. Thus:
%\vspace{-5pt}
\[
E = \eta_D \times \tau = \frac{2^{256}}{target} =
\frac{ 2^{256} }{target\_1}\times \frac{target\_1}{target} \approx 2\times \Delta 
\]
and conclude that
%\vspace{-5pt}
\begin{equation}
\tau = \frac {2 \times \Delta}{\eta_D}
\label{eq:time:to:hashrate:puzzler}
\end{equation}
\noindent
For instance, the lowest puzzle difficulty is 1, which occurs when the $target$
has a prefix of one zero and the device is expected to generate 2 hashes before
solving the puzzle.  Similarly, the maximum difficulty is $(2^{255}-1)$, for a
$target$ = 1, when the device is expected to perform $\frac{2^{255}-1}{1}
\times 2 \approx 2^{256}$ hashes.

%\ruben{This derivation also seems to suggest that the reported hashrate by
%clients is actually the double hash operation, corresponding to half the
%hashrate reported by openssl, for instance.}

\begin{figure}[t]
%\vspace{-10pt}
\renewcommand{\baselinestretch}{0.7}
\begin{minipage}{0.4\textwidth}
\begin{algorithm}[H]
\begin{tabbing}
XXX\=X\=X\=X\=X\=X\= \kill

1. \small{Object \mbox{\bf{FraudSysService}}}\\

2.\> \small{K: key;}\\

3. \> \small{Function\ {\bf BuildCookie}($U$, $D$, $S$, $A$, $q$)}\\

4.\>\> \small{$\eta_D$ := getHashrate($U, D$);}\\
5.\>\> \small{$r$ := computeFraudScore($U, S, A$);}\\
6.\>\> \small{$\tau$ := fraud2Penalty($r$);}\\
7.\>\> \small{$\Delta$ := $\eta_D \times \tau / 2q$}\\
8.\>\> \small{$oldT$ := getTimeout(U);}\\
9.\>\> \small{$newT$ := $oldT + \tau$;}\\
10.\>\> \small{$\Gamma$ := HMAC($K, U, D, S, newT, \Delta, A$);}\\
11.\>\> \small{setTimeout($U$, $newT$);}\\
12.\>\> \small{return $\Gamma$, $\Delta$, $newT$;}\\

13.\> \small{Function\ {\bf VerifyPuzzle}($U$, $D$, $S$, $A$, $timeout$, $\Gamma$, $\sigma$: share[$q$])}\\
14.\>\> \small{\mbox{\bf{if}}\ ($\Gamma$ != HMAC($K, U, D, S, A, timeout, \Delta$)\ return -1;}\\
15.\>\> \small{$target$ := getTarget($\Delta$);}\\
16.\>\> \small{\mbox{\bf{for}}\ ($i$ := 0;\ $i< q$;\ $i$++)}\\
17.\>\>\> \small{\mbox{\bf{if}}\ ($H^2(\sigma[i]\ ||\ \Gamma) > target$)\ return -1;}\\
18.\>\> \small{waitUntil($timeout$); post $A$;}\\
19.\>\> \small{$\tau' := T_c - T$;}\\
20.\>\> \small{\mbox{\bf{if}}\ (($\eta_D := 2 \Delta/\tau') \ge \eta_{min}$)}\\
21.\>\>\> \small{updateHashrate($U$, $D$, $\eta_D$);}\\

22. \small{Object\ \mbox{\bf{UserDevice}}}\\

23.\> \small{Function\ {\bf SolvePuzzle}($\Gamma$, $\Delta$, $timeout$, $q$)}\\
24.\>\> \small{$target$ := getTarget($\Delta$);}\\
25.\>\> \small{$\sigma$ := new share[q]; $i$ := 0;}\\
26.\>\> \small{\mbox{\bf{while}}\ ($i < q$)\ \mbox{\bf{do}}}\\
27.\>\>\> \small{$nonce$ := getRandom();}\\
28.\>\>\> \small{\mbox{\bf{if}}\ ($H^2(nonce\ ||\ \Gamma) < target$)}\\
29.\>\>\>\> \small{$\sigma[i]$ := $nonce$;}\\
30.\>\>\>\> \small{$i$ := $i$+1;}\\
31.\>\> \small{return\ $U, D, S, A, timeout, \Gamma, \sigma$;}

\end{tabbing}
\caption{FraudSys puzzle creation, verification and computation components.}
\label{alg:fraudsys}
\end{algorithm}
\end{minipage}
\normalsize
\vspace{-5pt}
\end{figure}

\vspace{0.05in}\noindent
{\bf The FraudSys puzzle and cookies}.
To minimize the storage imposed on the FraudSys service (see above), we
leverage the cookie concept~\cite{B96}. Algorithm~\ref{alg:fraudsys}
illustrates the puzzle creation, verification and computation components. The
FraudSys service generates and stores a secret key $K$ (line 2).  When a user
$U$ performs an activity $A$ from a device $D$ on a subject $S$ of the online service,
the online service calls the BuildCookie function of the FraudSys service
(lines 3-11).
BuildCookie retrieves the hashrate of the device $D$ from the record stored by
FraudSys for $U$ (line 4). It then computes the fraud score associated to the
activity (line 5) then converts it to a time penalty $\tau$ (line 6). We
describe this functionality in the next subsections. BuildCookie then uses a
modified Equation~\ref{eq:time:to:hashrate:puzzler} to compute the difficulty
$\Delta$ that the puzzle should have (line 7). $\Delta$ is $q$ times smaller
than in Equation~\ref{eq:time:to:hashrate:puzzler}, as the puzzle solution
consists of $q$ shares, see SolvePuzzle.
BuildCookie gets the current timeout $oldT$ of $U$, and updates it to $newT$ by
adding the penalty $\tau$ to it (lines 8-9). It then computes the puzzle cookie
$\Gamma$,
\[
\Gamma = HMAC_K (U, D, S, A, timeout, \Delta)
\]
as a keyed HMAC~\cite{BCK96} over the user and device id, subject, activity,
new timeout and puzzle difficulty (lines 9-10). BuildCookie sets $U$'s timeout
value to the updated $newT$ value (line 11), then returns the following puzzle
(line 12) to the online service that forwards it to device $D$ (see
Figure~\ref{fig:system}):
\[
\Pi = \Gamma, \Delta, timeout.
\]
The puzzle cookie ensures that an adversary that modifies the puzzle's
difficulty or timeout, will be detected: the adversary does not know the key
$K$, which is a secret of the FraudSys service. Puzzle cookies are unique with
high probability, due to collision resistance properties of the HMAC, whose
input is non-repeating.

\vspace{0.05in}\noindent
{\bf Solving the puzzle}.
When the device $D$ receives the puzzle, it needs to solve it: search for $q$
$nonce$ values that satisfy the inequality $H^2(nonce\ ||\ \Gamma) < target$,
for a $target$ corresponding to the difficulty $\Delta$. Specifically, $D$
invokes the SolvePuzzle function (lines 23-31), that needs to identify $q$ {\it
shares}, i.e., nonce values that satisfy the puzzle. $q$ is a system parameter.
The function first uses Equation~\ref{eq:delta:target} to retrieve the $target$
value corresponding to the difficulty $\Delta$ (line 24).  Then, it generates
random $nonce$ values until it identifies $q$ values that satisfy the puzzle
condition (lines 25-30). SolvePuzzle returns the identified shares (in the
$\sigma$ array), which are then sent to the online service and forwarded to the
FraudSys server, along with the user, device and subject ids, activity, timeout
and cookie of the received puzzle (see line 12 and Figure~\ref{fig:system}).

\begin{figure}
\centering
\includegraphics[width=0.47\textwidth]{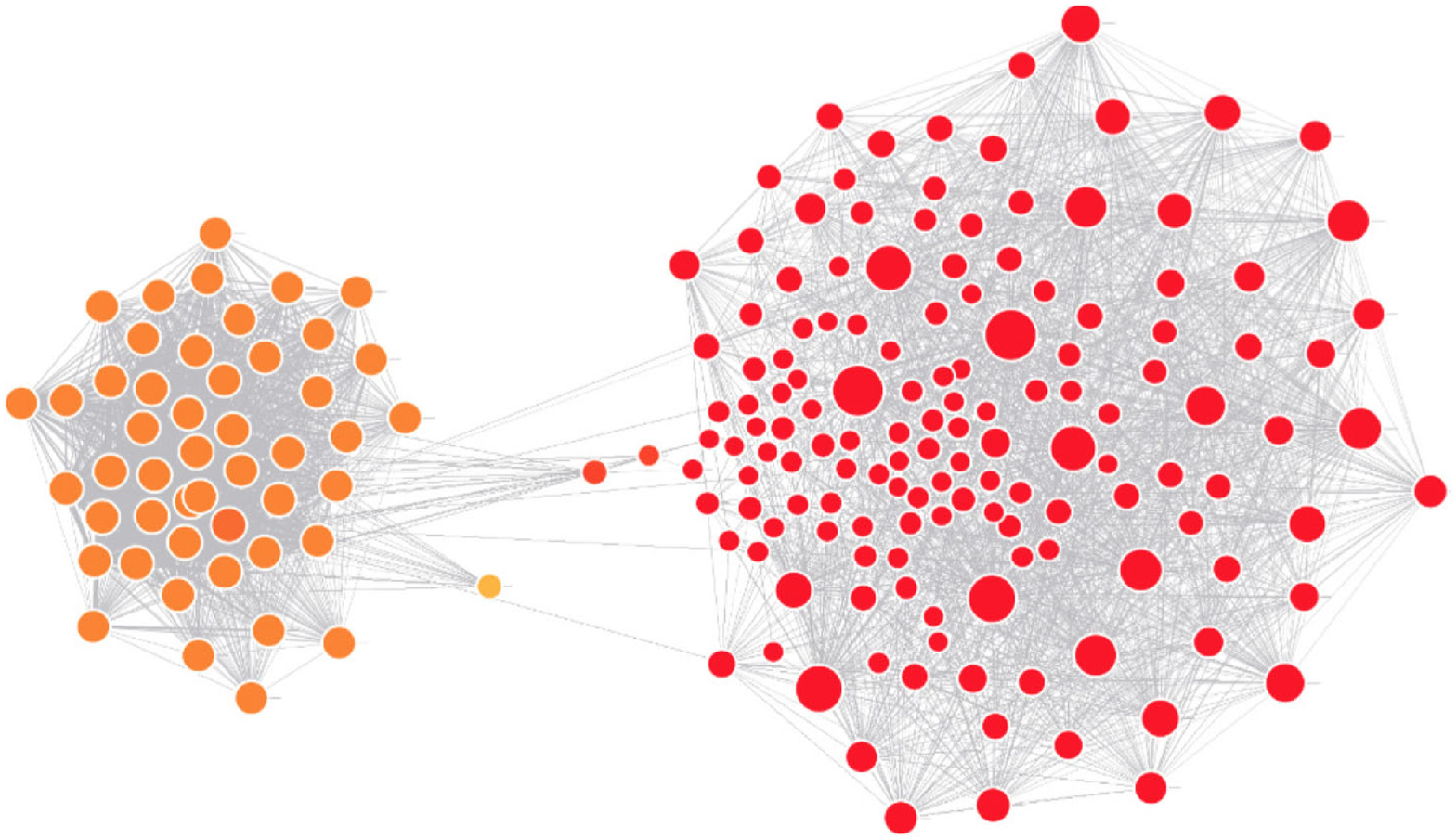}
\caption{Visualization of the co-review graph of a fraudulent Google Play app.
The nodes represent user accounts; edges connect nodes corresponding to
accounts with common, past review activities. The nodes in each of the 2
clusters correspond to accounts controlled by the same fraudster.}
\label{fig:coreview:fraud}
\vspace{-15pt}
\end{figure}

\vspace{0.05in}\noindent
\noindent
{\bf Verification of puzzle correctness}.
Upon receiving these values, the FraudSys server invokes the VerifyPuzzle
function (lines 13-21), to verify its correctness as follows:
%\item
%The current time does not exceed the timeout timestamp stored for $U$ (lines
%14-16). If it does, then the device has not been able to solve the puzzle in
%time. \bogdan{Ruben, is this really a problem? Why not accept the activity in
%this case and decrease the hashrate?}\ruben{agreed. We should have a lower
%bound though, to prevent an attacker to lower its hashrate below the hashrate
%of a nexus 4, for instance.} 
%$\bullet$
(1) Reconstruct the puzzle cookie $\Gamma$ based on the received values and the
secret key $K$.  Verify that this cookie is equal to the received $\Gamma$
value (line 14).  This ensures that all  values, including the $timeout$
have not been altered by an adversary; and 
(2)
%$\bullet$
Verify that each of the $q$ shares satisfy the puzzle (lines 15-17).
If these verifications succeed, FraudSys waits until $timeout$ expires to
confirm the user action $A$, for posting by the online service (line 18).
It then uses the time required by the device to solve the puzzle, to
re-evaluate the hashrate of the device (lines 19-20). It updates the stored
hashrate only if the new value is above a minimally accepted hashrate value
(lines 20-21).

\subsection{The Fraud Detection Module}
\label{sec:fraudsys:fraud}

To assign a fraud score to a user activity in real-time, the fraud detection
module can only rely on the existing history of the user and of the subject on
which the activity is performed. We propose an approach that builds on the {\it
co-activity graphs} of subjects, where nodes correspond to user accounts that
performed activities on the subject, and edges connect nodes whose user
accounts have a history of activities that targeted the same subjects. Edge
weights denote the size of that history. Figure~\ref{fig:coreview:fraud} shows
the co-review (where activities are reviews) graph of a fraudulent Google Play
app, that received fake reviews from 2 fraud workers. Each cluster is formed by
accounts controlled by one of the workers.

The fraud detection module leverages the adversary model findings
($\S$~\ref{sec:model:adversary}) that a fraudster-controlled user account that
performs a new activity on a subject, is likely to be well connected to the
co-activity graph of the subject, or at least one of its densely connected
sub-graphs.  Figure~\ref{fig:coaction:dynamic} illustrates this approach: Let
$U$ be a user account that performs an activity $A$ for a subject $S$ at time
$T$.  Let $G=(V, E)$ be the co-activity graph of $S$ before time $T$.  Let $G_T
= (V_T, E_T)$ be the new co-activity graph of $S$, that also includes $U$,
i.e., $V_T = V \cup U$. Given $U$, $S$ and $G$, FraudSys extracts the following
features, that model the relationship of $U$ with $S$:

\begin{figure}
\centering
\includegraphics[width=3.5in]{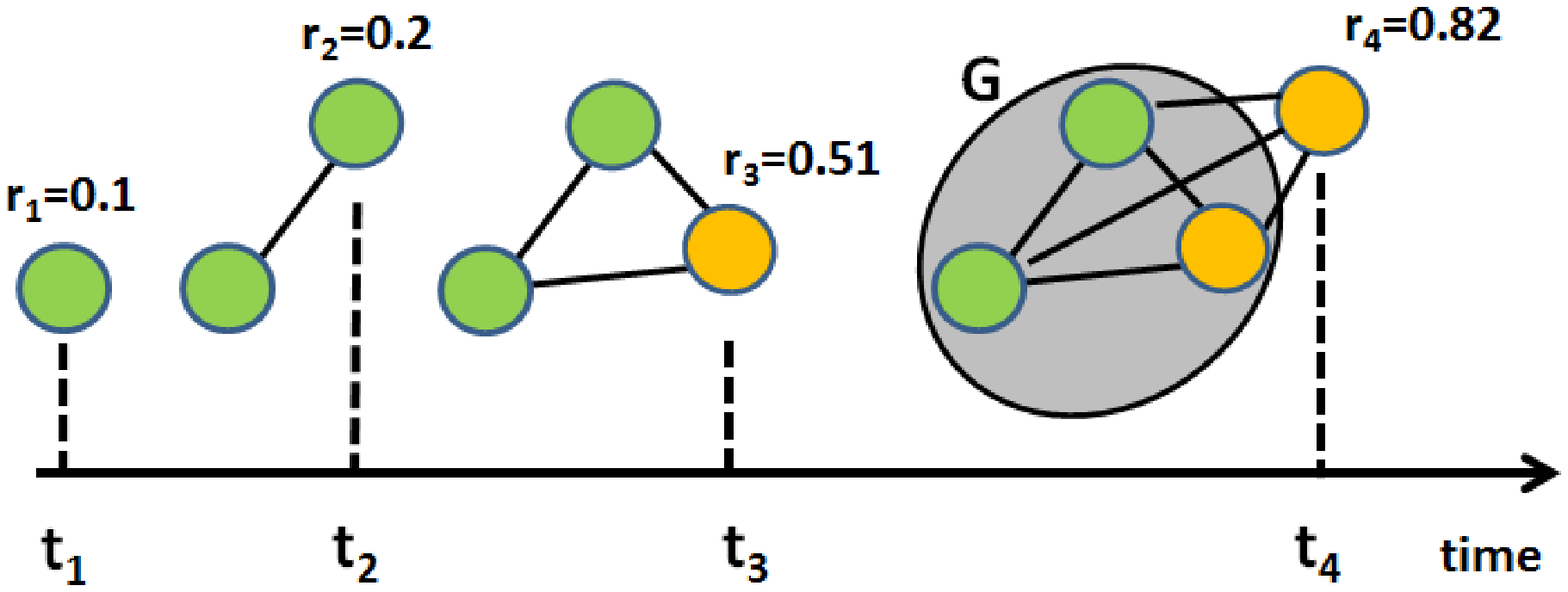}
\caption{Fraud detection illustration: temporal evolution of the co-activity
graph of a subject. The nodes represent user accounts that have performed an
activity on the subject. Edges connect accounts with common past activities. As
a new user account posts an activity, FraudSys assigns the activity a fraud
score (the $r_1 .. r_4$ values), based on its connectivity to previous
activities. Yellow nodes are considered fraudulent ($r >
0.5$).
\label{fig:coaction:dynamic}}
\vspace{-15pt}
\end{figure}

\vspace{0.05in}\noindent
$\bullet$
{\bf Connectivity features}.
The percentage of nodes in $V$ to whom $U$ is connected.  The average weight of
the edges between $U$ and the nodes in $V$. The average weight of those edges
divided by the average weight of the edges in $E$. This feature will indicate
if $U$ increases or decreases the overall connectivity of $G$. The number of
triangles in $G_T$ that have $U$ as a vertex. The average edge weight of those
triangles.

\vspace{0.05in}\noindent
$\bullet$
{\bf Best fit connectivity features}.
Since $U$ may be controlled by one of multiple fraudsters who target $S$, $U$
may be better connected to the subgraph of $G$ controlled by that fraudster.
Then, use a weighted min-cut algorithm to partition $G$ into components
$G_1,..,G_k$, such that any node in a component is more densely connected to
the nodes in the same component than to the nodes in any of the other
components.  $G_1,..,G_k$ may contain user accounts controlled by different
fraudsters, see Figure~\ref{fig:coreview:fraud}.  Identify the component $G_b$,
$b \in \{1,..k\}$ to which $U$ is the most tightly connected (according to
the above connectivity features).
%
%We conjecture that the fraudster who controls the accounts in $G_b$ is the
%most likely to also control $U$.
%
Output the connectivity features between $U$ and $G_b$.

\vspace{0.05in}\noindent
$\bullet$
{\bf Account based features}.
The number of activities previously performed by $U$.  The age of $U$: the time
between $U$'s creation and the time when activity $A$ is performed on $S$.  The
{\it expertise} of $U$: the number of actions of $U$ for subjects similar to
$S$. Similarity depends on the online service, e.g., same category apps in
Google Play, pages with similar topics in Facebook.

%\item
%Feature 5:
%the probabilities assigned to the neighbors of $U$ in the previous graph $G$.

\vspace{0.05in}\noindent
The Fraud Detection module trains a probabilistic supervised learning algorithm
on these features and uses the trained model to output the probability that a
given activity is fraudulent. We detail the performance of various algorithms,
over data that we collected from Google Play and Facebook, in
$\S$~\ref{sec:evaluation}.

\vspace{0.05in}\noindent
{\bf Per-fraudster timeout}.
We exploit the ability of the fraud detection module to identify accounts
controlled by the same fraudster, to further restrict fraud. Specifically,
instead of storing a $timeout$ timestamp for each user account, FraudSys can
store a single $timeout$ per detected fraudster. Thus, FraudSys will accumulate
penalties in a single, per-fraudster account. This facilitates
Claim~\ref{claim:fps}.

%\subsubsection{Discussion of Problems}
%
%1. Suppose a real user provides a review in app1 and then a fraudster provide
%100 reviews in the same app. In another app app2, the same fraudster provides
%100 reviews with the same accounts and then the same real user comes to give
%rating. This time he has to wait long to provide the review. We can solve this
%problem using a weight threshold: we ignore edges with small weights, as this
%user is likely poorly connected to the 100 fraudster accounts.
%
%2. A group of honest users could be interested in apps of a certain category,
%e.g., arcade games, and thus very likely to review many of those apps in
%common. We have observed this behavior among the apps whose data we have
%collected. One solution is to look at edge weights in more detail: instead
%of a single value, we could have multiple values, split among the categories
%of apps reviewed in common.

\subsection{The Fraud2Penalty Module}
\label{sec:fraudsys:all}

Given the fraud score $r$ of an activity of user $U$ (output by the Fraud
Detection module), performed from a device $D$ associated with the account of
$U$ (see the model section), the Puzzler module generates a puzzle whose
difficulty is a function of both $r$ and the computational capability of $D$.
We now describe the Fraud2Penalty module, that converts $r$ into a time
penalty.
\begin{figure}
\centering
\includegraphics[width=0.47\textwidth]{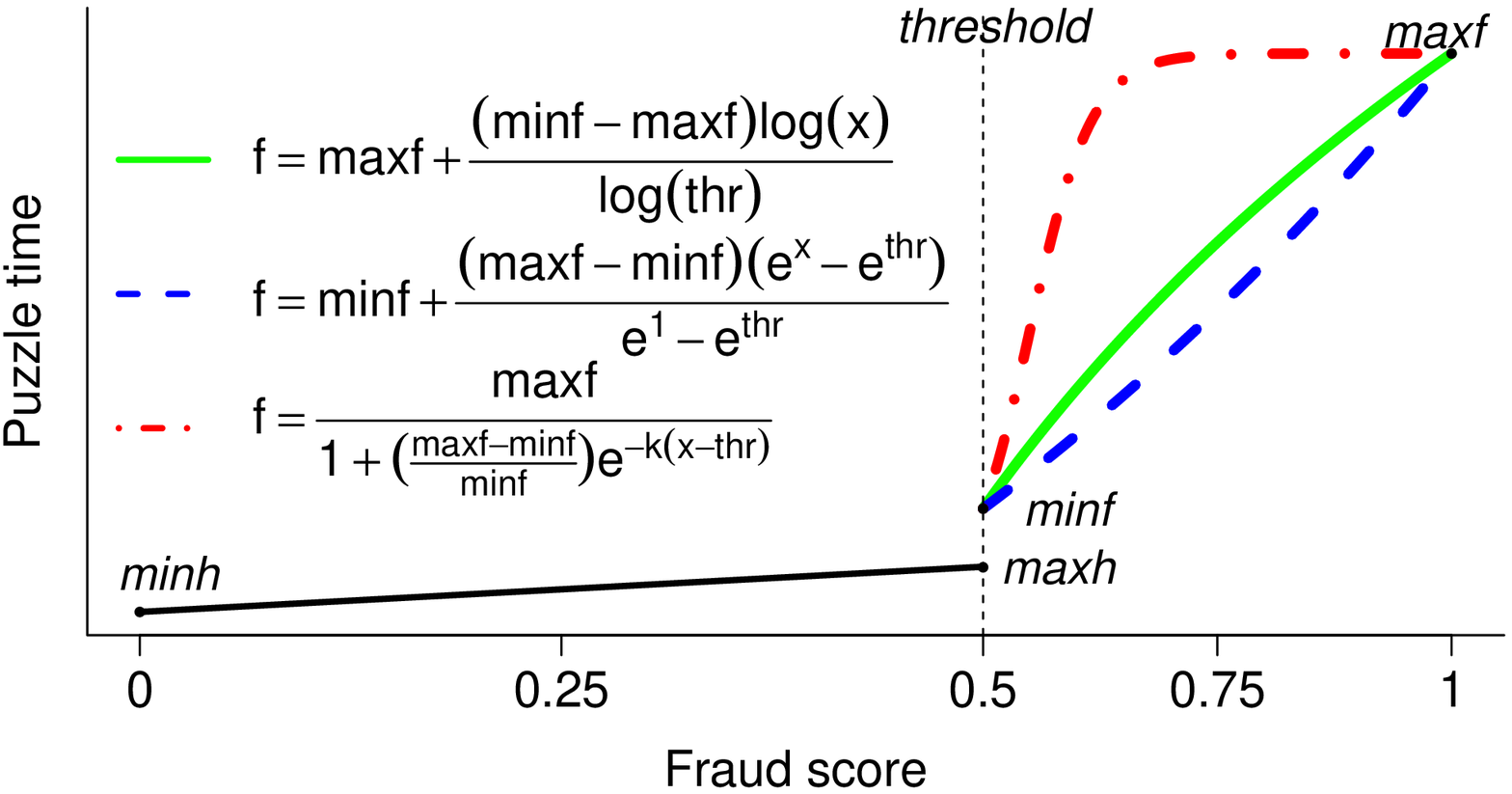}
\vspace{-5pt}
\caption{Comparison of functions to convert fraud scores (x axis) to time
penalties (y axis). The logistic function (red dot-line) exhibits the required
exponential increase.}
\label{fig:conversion}
\vspace{-15pt}
\end{figure}
We have explored several functions to convert the fraud score $r$ of a user
activity to a time penalty. Let $minh$ and $maxh$, and $minf$ and $maxf$,
denote the minimum and maximum times imposed on the device from which an
honest, respectively fraudulent activity is performed. Let $thr$ denote the threshold fraud score above which we start to consider a user
activity as being fraudulent. We propose a conversion function that increases
linearly when $r < thr$, and exponentially when $r > thr$. Specifically, we propose a flexible generalization of
the logistic function (when $r > thr$), where the parameter
$k$ is the {\it growth rate}:
\begin{equation}
\begin{cases}
\frac{maxh - minh}{thr} r + minh  & 0 \leq r \leq thr \\
\frac{maxf}{1 + (\frac{maxf-minf}{minf})e^{-k(r-thr)}} & thr \leq r \leq 1
\end{cases}
\label{eq:fraud2penalty}
\end{equation}
\noindent
We have compared this logistic increase function with other functions, with the
same linear increase in the honest region, but exponential ($(maxf -
minf)\frac{e^{r} - e^{thr}}{e^{1} - e^{thr}}
+ minf$) and logarithmic ($(minf - maxf)\frac{log r}{log(thr)}
+ maxf$) increase in the fraudulent regions.  Figure~\ref{fig:conversion}
compares the logistic, exponential and logarithmic functions. It shows that
unlike the exponential and logarithmic functions, the logistic function
exhibits the desired rapid increase for fraud probability values above the
threshold value.
In $\S$~\ref{sec:evaluation} we detail parameter values for the logistic
conversion function,

\subsection{The Hashrate Inference Module}
\label{sec:fraudsys:hashrate}

\noindent
{\bf New device registration}.
When a user registers a new device, the device sends its specs to the online
service that forwards them to FraudSys. FraudSys leverages its
list of profiled devices (see Table~\ref{table:hashrate}) to retrieve the
hashrate of the profiled device with the most similar capabilities. FraudSys
stores the new device along with this initial hashrate estimate under the id of
the user that registers it (see the Puzzle module).
Given this hashrate and the
above time penalty, FraudSys uses Equation~\ref{eq:time:to:hashrate:puzzler}
to compute an initial puzzle difficulty.

\vspace{0.05in}\noindent
{\bf Hashrate correction}.
The initial hashrate estimate of FraudSys may be incorrect. In addition, as
discussed in the System Model, the user may be adversarial, thus attempt to
provide an inaccurate view of the puzzle solving capabilities of his device.
To address these problems, FraudSys employs an adaptive hashrate correction
process. Specifically, an adversary with a more capable device than advertised
(see e.g., Table~\ref{table:hashrate}) will solve the assigned puzzle faster.
The incentive for this is a shorter wait time for his activity to post on the
online service. If this occurs, FraudSys increases its device hashrate
estimate to reflect the observed shorter time required by the device to solve
the puzzle (see Algorithm~\ref{alg:fraudsys}, lines 19-20).

\section{FraudSys Properties}
\label{sec:properties}

\begin{claim}
A fraudster that performs a fraudulent activity with fraud score $r$ from a
device with hashrate $\eta$, is expected to compute $\frac{\eta \times maxf}{1
+ (\frac{maxf-minf}{minf})e^{-k(r-thr)}}$ double hashes.
\label{claim:work}
\end{claim}

\vspace{-12pt}

\begin{proof}
According to Equation~\ref{eq:fraud2penalty}, the time penalty assigned to a
fraudulent activity with score $r$ is $\tau = \frac{maxf}{1 +
(\frac{maxf-minf}{minf})e^{-k(r-thr)}}$. Then,
Equation~\ref{eq:time:to:hashrate:puzzler} ensures that the number of expected
hashes that the device needs to perform to solve the puzzle of
Equation~\ref{eq:puzzle} is $\eta \times \tau$, which concludes the proof.
\end{proof}

\vspace{-5pt}

Let $f$ be the number of fraud workers in the system (i.e., $f =
|\mathcal{F}|$), $\tau$ be the average temporal penalty assigned by FraudSys
to a fraudulent activity, and let $p$ be the expected payout for a single
fraudulent activity. We introduce then the following claim:

\vspace{-5pt}

\begin{claim}
FraudSys is a $(p/\tau, f/\tau)$-fraud preemption\\ system.
\label{claim:fps}
\end{claim}

\vspace{-13pt}

\begin{proof}
The best fit connectivity features of the Fraud Detection module (see
$\S$~\ref{sec:fraudsys:fraud}) enable FraudSys to detect activities performed
from accounts controlled by the same fraudster.  This, coupled with an
extension of the $timeout$ concept applied at the fraudster level (see
$\S$~\ref{sec:fraudsys:fraud}) ensures a serialization of fraudster activities.
Then, the average number of fraudulent activities that a fraudster can post per
time unit in FraudSys is $1/\tau$. This implies that, per time unit, the
expected payout of a fraudster is $p/\tau$, and a subject can be the target of
at most $f/\tau$ fraudulent activities. This, according to the definition of
$\S$~\ref{sec:model:problem}, completes the proof.
\end{proof}

\vspace{-5pt}

\subsection{Security Discussion and Limitations}

The FraudSys puzzle not only ties the penalty computation to the user
activity, but also addresses pre-computation, replay and guessing attacks: the
adversary cannot predict the cookie value of its actions, thus cannot
pre-compute puzzles and cannot reuse old cookies. It also offloads significant
work from the FraudSys service, which no longer needs to keep track of puzzle
assignments.

%The user activity is self-contained with its puzzle and proof-of-work.

\vspace{0.05in}\noindent
{\bf Device deception}.
An adversary with a specialized puzzle solving device (e.g., AntMiner) will be
assigned puzzles with large difficulty values (see, e.g.,
Table~\ref{table:hashrate}), thus consume the same amount of time as when using
a resource constrained device (e.g., a smartphone). The adversary can exploit
this observation to avoid the implications of Claim~\ref{claim:work}: register
a resource constrained device, but rely on a powerful back-end device to solve
the assigned puzzles faster. The adversary has two options. First, report the
solutions as soon as the back-end device retrieves them. In this case however,
the adversary leaks his true capabilities, as FraudSys will update the
adversary hashrate (Algorithm~\ref{alg:fraudsys}, line 20). Thus,
subsequently, his assigned puzzles will have a significantly higher difficulty
value. In a second strategy, the adversary estimates the time that his
front-end device would take to complete the puzzle, then waits the remaining
penalty time. In this case, the adversary incurs two penalties, the long
wait time and the underutilized back-end device investment.

%\noindent
%{\bf Resource constrained devices}.
%Instead of investing in expensive mining equipment, the adversary may choose to
%use a resource constrained device, e.g., old smartphones. FraudSys will assign
%such devices relatively easy puzzles, as their hashrate is likely low.  The
%adversary then has two options. First, let the device perform large amounts of
%computation. \bogdan{Find here data on burnout rate of mobile devices when
%exposed to mining computations.} Second, the adversary can use the mobile
%device as a front-end to a more computation intensive back-end (e.g., and
%AntMiner device): the smartphone receives the puzzle, sends it to the back-end.
%This is equivalent to the device pretender strategy, see above.

\vspace{0.05in}\noindent
{\bf Adversary strategies: new user accounts}.
To avoid the implications of Claim~\ref{claim:fps}, the adversary registers new
user accounts. While new accounts are cheap, their freshness and lack of
history will enable the account based features of the Fraud Detection module to
label them as being likely fraudulent. As the adversary reuses such accounts,
the connectivity features start to play a more important role in labeling their
activities as fraudulent. Thus, the adversary has a small usable window of
small penalties for new accounts.

While new honest accounts may also be assigned larger penalties for their first
few activities, they will not affect the user experience: the user can continue
her online activities, while her device solves the assigned puzzle in the
background.

\section{Empirical Evaluation}
\label{sec:evaluation}

%\mizan{To extract the features we have used Java. To For data processing and
%penalty calculation we have used python, to calculate probabilty we have used
%scikit learn which is also a python module. To plot data we have used R.}

%In this section we first describe the data we have used and our efforts to
%profile the puzzle solving capabilities of various devices, then evaluate the
%ability of FraudSys to accurately detect and effectively penalize fraudulent
%behaviors on the collected data.

\subsection{Datasets}
\label{sec:data}

\begin{figure*}
\centering
\captionsetup[subfigure]{aboveskip=0.1in, belowskip=-0.05in}
\begin{subfigure}[b]{2.25in}
\includegraphics[width=2.25in]{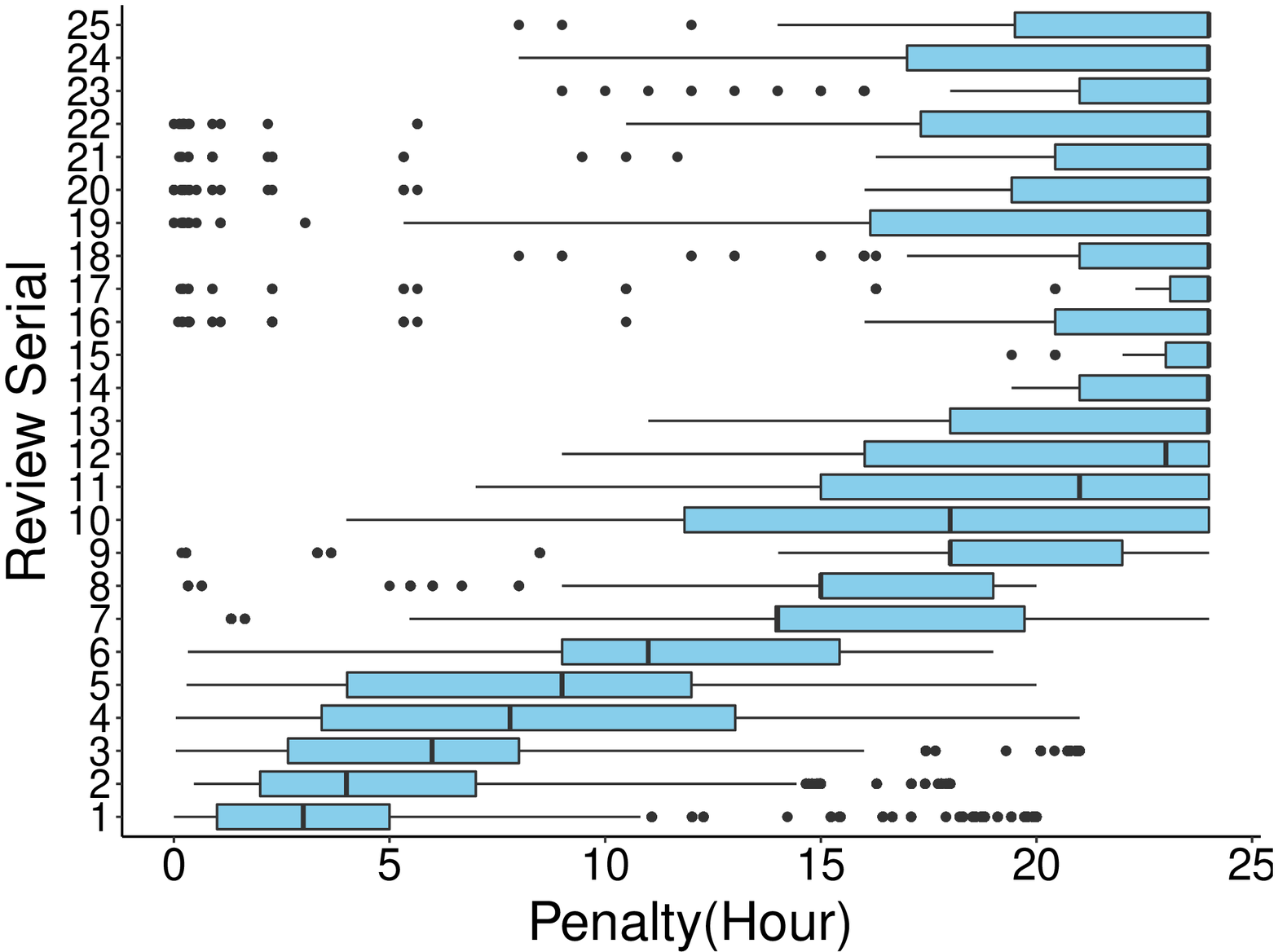}
\caption{}
\label{fig:penalty:evolution}
\end{subfigure}
~
\captionsetup[subfigure]{aboveskip=0.1in, belowskip=-0.05in}
\begin{subfigure}[b]{2.25in}
\includegraphics[width=2.25in]{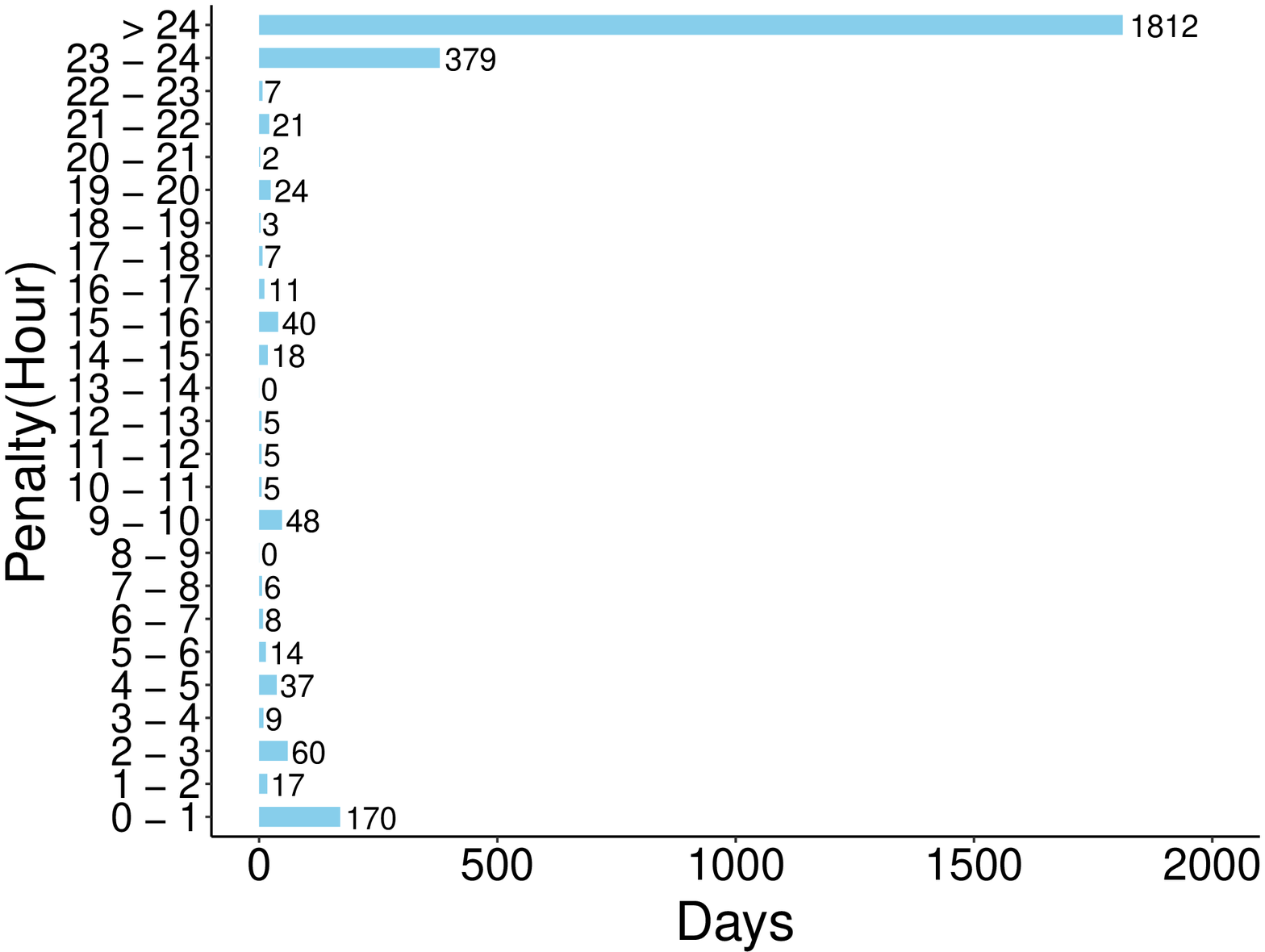}
\caption{}
\label{fig:gplay:fraud}
\end{subfigure}
~
\begin{subfigure}[b]{2.25in}
\includegraphics[width=2.25in]{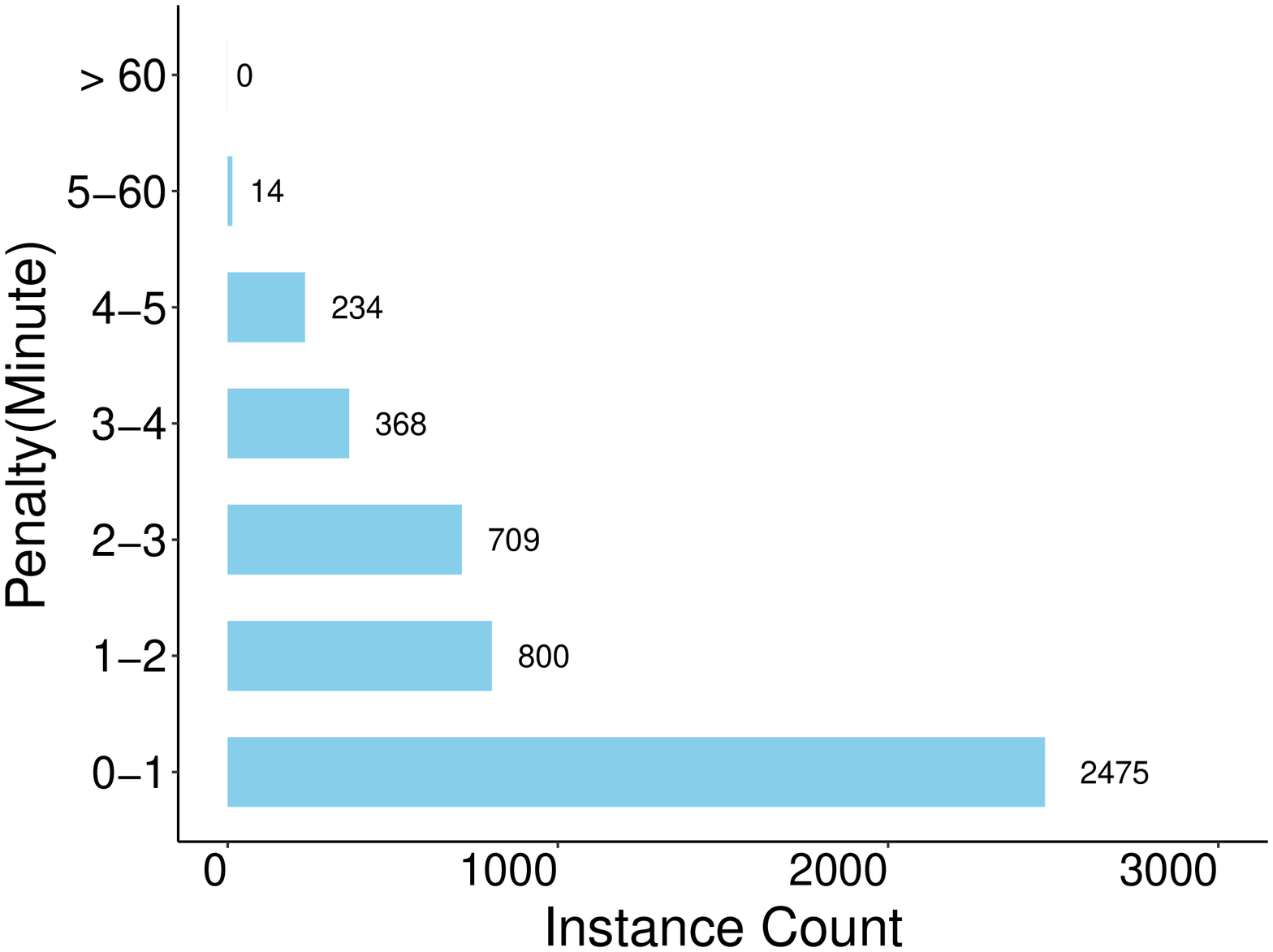}
\caption{}
\label{fig:gplay:honest}
\end{subfigure}
\vspace{-5pt}
\caption{
Stats over the Google Play data when $maxf$ = 24h, $minh$ = 2s, $maxh = minf$ = 5min.
(a) Evolution of average, 1st and 3rd quartile of the penalty imposed on the
$i$-th fraud activity of a fraudster for the same subject. It
shows a steep increase: the average penalty of the first three fraud activities
for a subject sums to 15.34h, while {\bf the average penalty of the $12$th activity
exceeds 24h}.
(b) Distribution of per-fraudster daily penalties, over data from 23
fraudsters: in 1,812 days out of 2,708 days, the penalty assigned to a single
fraudster exceeds 24 hours.
(c) Distribution of penalties assigned to an honest review. Only 14 out of
4,600 honest review instances received a penalty exceeding 5 minutes, but still
below 1 hour.}
\label{fig:gplay}
\vspace{-15pt}
\end{figure*}

We have collected the following datasets of fraudulent and honest behaviors
from Google Play and Facebook.

\vspace{0.05in}\noindent
{\bf Google Play: fraud behavior data}.
We have identified 23 workers in Freelancer, Fiverr and Upwork, with proven
expertise on performing fraud on Google Play apps.  We have
contacted these workers and collected the ids of 2,664 Google Play accounts
controlled by them.  We have also collected 640 apps heavily reviewed from
those accounts, with between 7 and 3,889 reviews, of which between 2\% and
100\% (median of 50\%) were written from accounts controlled by the workers.
These apps form our gold standard {\it fraud app} dataset.  We have also
collected the 23,028 fake reviews written from the 2,664 fraudster controlled
accounts for the 640 apps.  Figure~\ref{fig:coreview:fraud} shows the co-review
graph of one of these apps, that received fake reviews from 2 of the identified
23 workers.

%Using those 2,664 fraud users we have retrieved 39117 fraud reviews from 640
%fraud apps.

\vspace{0.05in}\noindent
{\bf Google Play: honest behavior data}.
We have selected $925$ candidate apps that have been developed by Google
designated ``top developers''. We have removed the apps whose apks
(executables) were flagged as malware by VirusTotal. We have manually
investigated $601$ of the remaining apps, and selected a set of $200$ apps that
(i) have more than $10$ reviews and (ii) were developed by reputable media
outlets (e.g., NBC, PBS) or have an associated business model (e.g., fitness
trackers). We call these the gold standard {\it benign app dataset}.

We have identified 600 reviewers of these 200 benign apps and 140 reviewers of
the 640 fraud apps (see above), such that each has reviewed at least 10 paid
apps, i.e., paid to install the app, then reviewed it, and had at least 5 posts
on their associated Google Plus (social network) accounts. These 740 user
accounts form our gold standard {\it honest user dataset}. We have then
retrieved and manually vetted 854 reviews written by the 600 honest reviewers
for the 200 benign apps, and 207 reviews written by the 140 honest reviewers of
the 640 fraud apps. Each selected review is informative, containing both
positive and negative sentiment statements. We call the resulting dataset, the
{\it honest review dataset}, with 1,061 reviews.

\vspace{0.05in}\noindent
{\bf Facebook Like dataset}.
We have used a subset of the dataset from~\cite{BLLTZ16}, consisting of 15,694
Facebook pages, that each has received at least 30 likes. The pages were liked
from 13,147 user accounts, of which 6,895 are fraudster controlled, and 6,252
are honest. In total, these fraudsters have posted 274,297 fake likes, and the
honest accounts have posted 180,400 honest likes.

\subsection{Device Hashrate Profile}

\begin{table}
\setlength{\tabcolsep}{.16677em}
\centering
\small
\textsf{
\begin{tabular}{l | r | r r r}
\toprule
\textbf{Device} & \textbf{Hashrate} & \textbf{Diff (5s)} & \textbf{ (12hr)} & \textbf{ (7 day)}\\
\midrule
Nexus 4 & 6.53 KH/s& 16.32K& 141.04M& 1.97G\\
Nexus 5 & 13.26 KH/s & 33.15K& 286.41M& 4.00G\\
LG Leon LTE & 10.1 KH/s& 25.25K& 218.16M& 3.05G\\
%FirePro 5950& 96.7 MH/s & 241.75M& 2.08T& 29.24T\\
NVS 295 & 1.7MH/s & 4.25M& 36.72G& 514.08G\\
Server & 80 MH/s & 200M & 1.72T & 24.19T\\
AntMiner & 4.72 TH/s & 11.8T& 101.95P& 1427P\\
\bottomrule
\end{tabular}
}
\caption{Hashrate profiling table for various device types (smartphone, tablet,
PC and Bitcoin miner), along with difficulty values for penalty times of 5s, 12
hours and 7 days.}
\label{table:hashrate}
\vspace{-20pt}
\end{table}

\begin{table}[h]
\vspace{-5pt}
\centering
\textsf{
\begin{tabular}{l | r r | r}
\toprule
\textbf{Strategy} & \textbf{FPR}\% & \textbf{FNR}\% & \textbf{Accuracy}\%\\
\midrule
%Bagging & 94.29 & 3.4 & 5.71 & 95.57\\
k-NN  & 1.41 & 4.45 & 97.92\\
SVM  & 5.8 & 11.3 & 92.40\\
Random Forest  & 3.44 & 6.46 & 95.69\\
\bottomrule
\end{tabular}
}
\caption{10-fold cross validation results of supervised learning algorithms in
fraud vs. honest Google Play review classification. k-NN achieves the lowest
FPR and FNR.}
\label{table:fraudsys:algorithms}
\vspace{-15pt}
\end{table}

We have profiled the hashrate of several devices, ranging from smartphones to a
Bitcoin mining hardware (AntMiner S7: ARMv7 CPU, 254 Mb of RAM, 135 BM1385
chips @ 700MHz).  Since Bitcoin mining requires capabilities far exceeding
those of smartphones, we have implemented an Android app to evaluate the
hashrate of several Android devices.
Table~\ref{table:hashrate} shows the hashrate values for the profiled devices,
along with the corresponding difficulty ($\Delta$) values for puzzles required
to impose 5 second, 12 hour and 7 day time penalties on such devices. We
observe the significant gap between the hashrate of a smartphone (10-15 KH/s)
and a specialized device (4.72 TH/s). This motivates the need for the puzzles
issued by FraudSys to have different $\Delta$ values for various user devices.
FraudSys maintains a similar table in order to be able to build appropriate
puzzles for newly registered devices.

%\vspace{-0.05in}
\subsection{Fraud Penalty Evaluation: Google Play}

\begin{figure*}
\centering
\captionsetup[subfigure]{aboveskip=0.1in, belowskip=-0.05in}
\begin{subfigure}[b]{2.25in}
\includegraphics[width=2.25in]{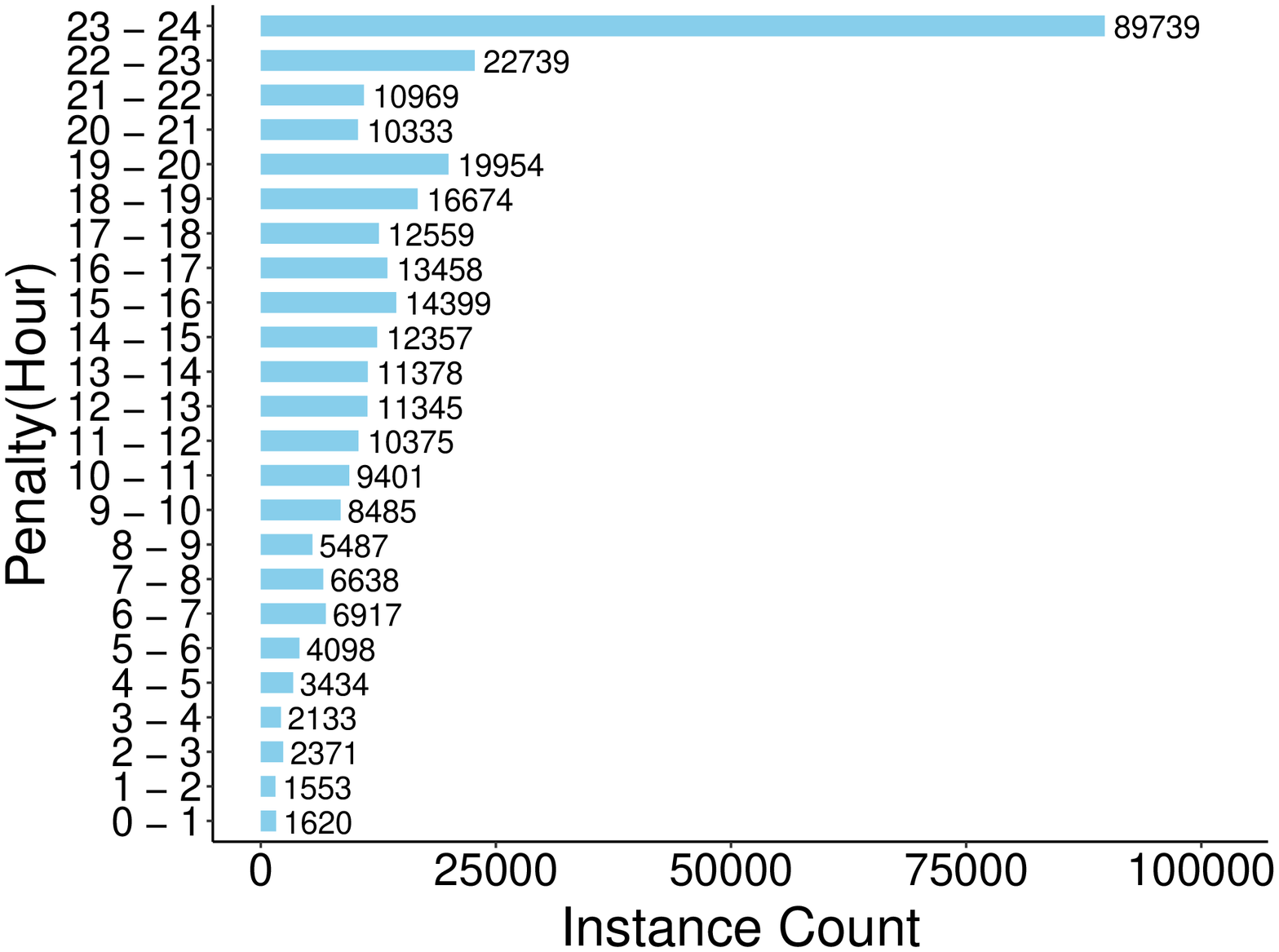}
\caption{}
\label{fig:facebook:fraud:penalty}
\end{subfigure}
~
\captionsetup[subfigure]{aboveskip=0.1in, belowskip=-0.05in}
\begin{subfigure}[b]{2.25in}
\includegraphics[width=2.25in]{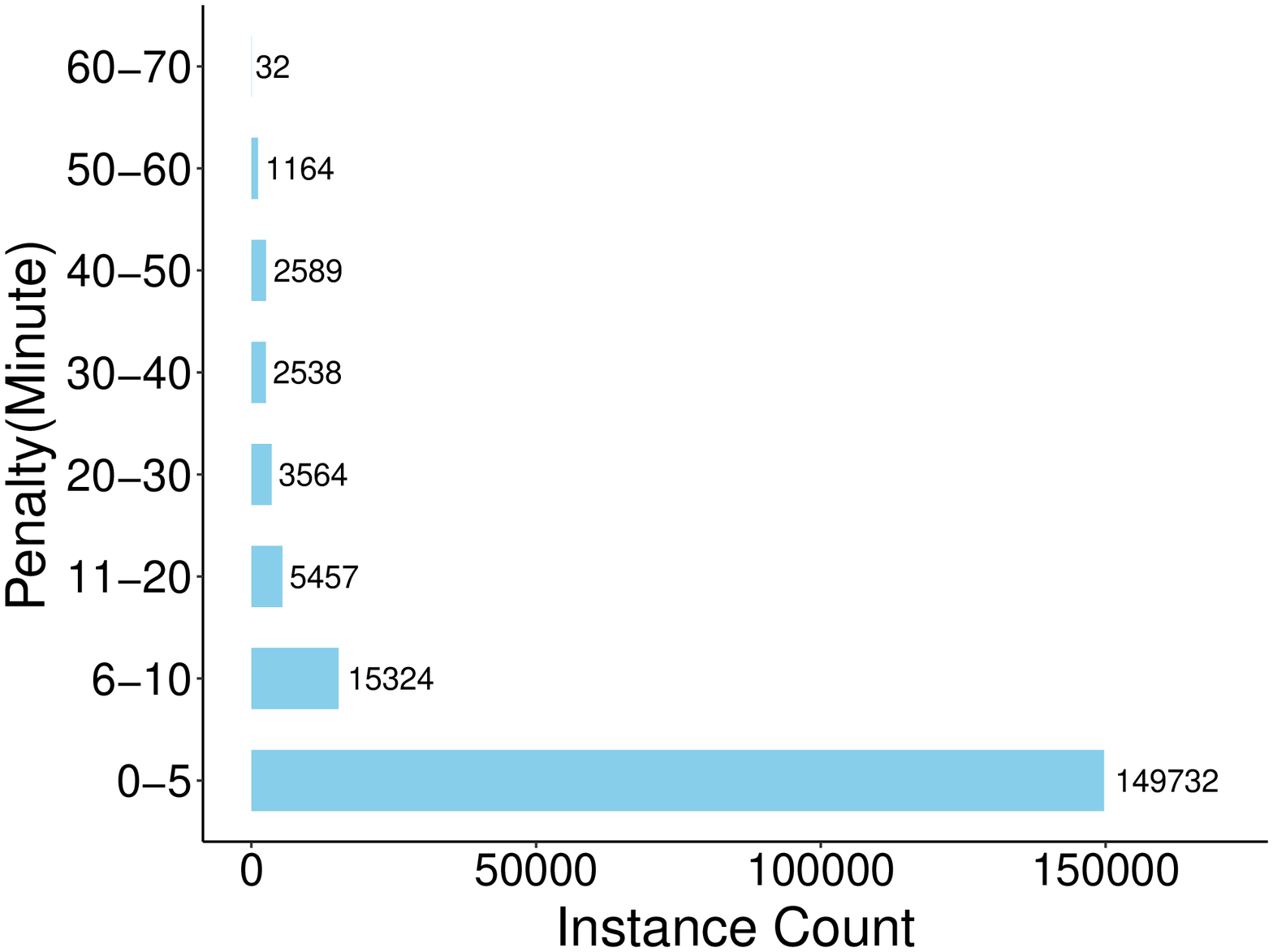}
\caption{}
\label{fig:facebook:honest:penalty}
\end{subfigure}
~
\begin{subfigure}[b]{2.25in}
\includegraphics[width=2.25in]{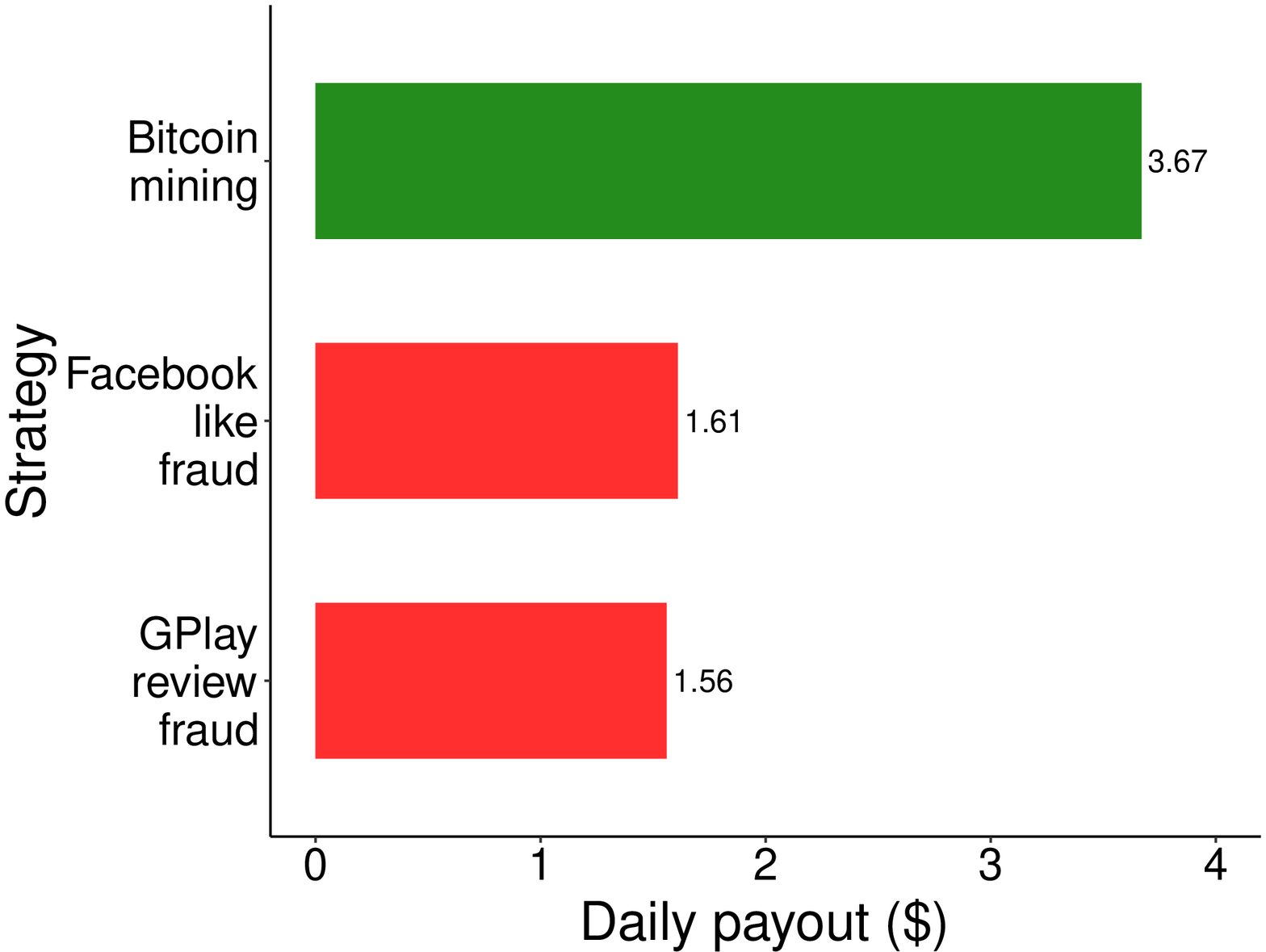}
\caption{}
\label{fig:payouts}
\end{subfigure}
\vspace{-5pt}
\caption{
(a) Penalty distribution for the fake Facebook likes. 84\% of the likes
received a penalty that exceeds 12 hours, and the average fake like penalty is
19.32 hours.
(b) Penalty distribution for the honest Facebook likes. 82.97\% of the honest
likes are assigned a penalty of under 5 min. The maximum penalty assigned
to an honest like is 70 minutes.
(c) Comparison of daily payouts provided by Bitcoin mining, writing fake
reviews in Google Play and posting fake likes in Facebook, under FraudSys.
{\bf Fraud does not pay off under FraudSys}: the fraud payout is less than half
the Bitcoin mining payout.}
\label{fig:facebook}
\vspace{-5pt}
\end{figure*}

\noindent
{\bf Supervised learning algorithm choice}.
We first used 10 fold cross-validation to evaluate the ability of the Fraud
Detection module to correctly classify the 23,028 fraudulent vs. 1,061 honest
reviews of the Google Play dataset previously described.
Table~\ref{table:fraudsys:algorithms} shows the false positive (FPR) and
negative (FNR) rates, as well as the accuracy achieved by the top 3 performing
supervised algorithms. k-NN has the lowest FPR and FNR, for an accuracy of
97.92\%. Thus, in the following experiments we use only k-NN.

\vspace{0.05in}\noindent
{\bf Parameter evaluation}.
We have used the fraud and honest review datasets described earlier, to compute
the temporal penalties imposed by FraudSys on fraudsters and honest users. We
have performed the following experiments. In each experiment, we use the data
of 22 fraud workers and 200 randomly chosen honest reviews (out of 1,061) to
train the supervised learning algorithm (k-NN) then test the model on the data
of the remaining fraud worker and on the remaining 861 honest reviews.  Thus,
we have performed 23 experiments, one for each worker.

We set the $maxf$ parameter such that the average daily payout of a fraudster
is below the average Bitcoin mining payout with a last generation AntMiner
device. Thus, this ensures that even such a powerful adversary has more
incentive to do Bitcoin mining instead of search rank fraud.  Specifically, the
above AntMiner's current (Jan. 2017) average daily payout is 0.0037 BTC. At
the current BTC to USD rate, this means \$3.67 per day~\footnote{Historically
speaking, the BTC to USD rate is increasing. The next generation AntMiner
coming up this year is expected to be 3 times more capable.}.  In addition, we
have experimented with $maxf$ values ranging from 12 to 48 hours.  The average
penalty assigned by FraudSys to a fraudulent review is 8.01 hours when
$maxf$=12h, 15.34h when $maxf$=24h, and 29.33h when $maxf$=48h.
Figure~\ref{fig:penalty:evolution} shows the median, first and third quartiles
for the time penalty (in hours) imposed on the $i$-th fraudulent activity
performed by a fraudster for a subject, when $maxf$= 24h: the 12th fake
activity receives a median penalty of 24h.

Thus, we set $maxf$=24h, which is sufficient for Google Play reviews: A
fraudster would be able to post on average less than 2 fake reviews per day,
thus, even with a reward of \$2 per fraud activity (see
Figure~\ref{fig:fraudsites}), achieve a payout of around \$3.15 per day, below
the Bitcoin mining payout. In addition, we have set $minh$ = 2s.
Figure~\ref{fig:timeline} shows the penalty timelines of two workers when
$minh$ = 2s, $maxh = minf$ = 5 min, $maxf$ = 24 hours, $thr = 0.5$, and $k=30$
(for a steep increase of time penalty with fraud score).  We note that a $maxh$
= 5 min is not excessive: this penalty is not imposed on the user, but on his
device. The user experience remains the same in the online service.

Each vertical bar shows the daily temporal penalty assigned to a single worker,
over reviews posted from multiple accounts. The maximum daily penalty of the
two workers is 1,247 hours and 3,079 hours respectively. We observe that each
worker has many days with a daily penalty exceeding 24 hours.

Figure~\ref{fig:gplay:fraud} shows for $maxh = minf$ = 5min, the overall
distribution of daily penalties assigned by FraudSys, over all the 23 fraud
workers, in the above experiment. It shows that during most of the active days,
fraud workers are assigned a daily penalty exceeding 24 hours.
Figure~\ref{fig:gplay:honest} (also for $maxh = minf$ = 5min) shows the
distribution of per-review penalty assigned by FraudSys to honest reviews,
shown over 4,600 (23 $\times$ 200) honest reviews. Irrespective of the $maxh$
value, only 14 honest reviews were classified as fraudulent, but assigned a
penalty below 1 hour. We observed minimal changes in the distribution of
penalties of fraudulent reviews when $maxh = minf$ ranges from 5 to 15 minutes.

%Since 99.39\% of the honest reviews are a penalty below 5min, in the following
%we set $maxh = minf$ = 5min.

\subsection{Fraud Penalty Evaluation: Facebook}

We have performed a similar parameter analysis using the Facebook ``like''
dataset.  Since this dataset lacks information about the fraudsters who control
the accounts that posted fake likes, we focus on the penalties assigned by
FraudSys to fake and honest likes.

Figure~\ref{fig:facebook:fraud:penalty} shows the distribution of penalties
assigned to fake likes and Figure~\ref{fig:facebook:honest:penalty} shows the
distribution of the honest likes.  Compared to the results over the Google Play
data, we observe a higher FPR, i.e., more honest likes with fraud level
penalties. We posit that this is due to the fewer features that we can extract
for the Facebook likes, as, unlike for Google Play reviews, we lack the time of
the activity. Specifically, absence of like sequence information enables us to
only extract features based on the last ``snapshot'' of the page, and not the
current page snapshot when the like was posted.

However, 82.97\% of the honest likes receive a penalty of under 5 mins and the
maximum penalty assigned to an honest review is 70 mins. In addition, 84\% of
the fake likes receive a penalty that exceeds 12 hours, and the average penalty
for a fake like is 19.32 hours. Figure~\ref{fig:payouts} compares the daily
payouts received by an AntMiner equipped fraudster who writes fake reviews in
Google Play (at \$1 per fake review), posts fake likes (at \$2 per fake like),
or honestly uses his device to mine Bitcoins. It shows that under FraudSys,
fraud doesn't pay off: the Bitcoin mining payout is more than double the fraud
payout for either fake reviews or likes.

\section{Conclusion}

We have introduced the concept of real-time fraud preemption systems, named as the FraudSys, that seek
to restrict the profitability and impact of fraud in online systems.  We
propose and develop stateless, verifiable computational puzzles, that impose
minimal overheads, but enable their efficient verification.  We have developed
a graph based, real-time algorithm to assign fraud scores to user activities
and mechanisms to convert scores to puzzle difficulty values. We used data
collected from Google Play and Facebook to show that our solutions impose
significant penalties on fraudsters, and make fraud less productive than
Bitcoin mining.

\section{Acknowledgments}

This research was supported in part by NSF grants CNS-1527153, SES-1450619,
CNS-1422215, IUSE-1525601, and Samsung GRO.

%\vspace{-0.1in}
\bibliographystyle{abbrv}
\bibliography{bitcoin,bogdan,crowdsource,fraud,graph,reviews,social.fraud}

\end{document}